# Designing hybrid graphene oxide- gold nanoparticles for nonlinear optical response: Experiment and theory

Rajesh Kumar Yadav[1], J. Aneesh[1], Rituraj Sharma[1], P. Abhiramnath[1], Tuhin Kumar Maji[2], Ganesh Ji Omar[1], A. K. Mishra[1], Debjani Karmakar[3]*and K. V. Adarsh[1]*

[1]Department of Physics, Indian Institute of Science Education and Research, Bhopal 462066 India

[2]Department of Chemical, Biological and Macromolecular Sciences, S. N. Bose National Centre for Basic Sciences, Salt Lake, Kolkata 700 098, India

[3]Technical Physics Division, Bhabha Atomic Research Centre, Mumbai-400085, India

Nonlinear optical absorption of light by materials are weak due to its perturbative nature, although a strong nonlinear response is of crucial importance to applications in optical limiting and switching. Here we demonstrate experimentally and theoretically an extremely efficient scheme of excited state absorption by charge transfer between donor and acceptor materials as the new method to enhance the nonlinear absorption by orders of magnitude. With this idea, we have demonstrated strong excited state absorption (ESA) in reduced graphene oxide that otherwise shows increased transparency at high fluence and enhancement of ESA by one orders of magnitude in graphene oxide by attaching gold nanoparticles (AuNP) in the tandem configuration that acts as an efficient charge transfer pair when excited at the plasmonic wavelength. To explain the unprecedented enhancement, we have developed a five-level rate equation model based on the charge transfer between the two materials and numerically simulated the results. To understand the correlation of interfacial charge-transfer with the concentration and type of the functional ligands attached to the graphene oxide sheet, we have investigated the AuNP-graphene oxide interface with various possible ligand configurations from first-principles calculations. By using the strong ESA of our hybrid materials, we have fabricated liquid cell-based high-performance optical limiters with important device parameters better than that of the benchmark optical limiters.

*Authors to whom correspondence should be addressed; debjan@barc.gov.in and adarsh@iiserb.ac.in

# I. INTRODUCTION

Strong nonlinear optical absorption of photons by materials is a long-standing goal of both fundamental and technological interest, for instance, critical for optical limiting and switching applications [1-4]. The ideal materials for optical limiting applications must have high transmission at low ambient intensity, reduced transmission at high intensity and a large dynamic range at which the device damages (irreversibly) to the limiting input. Optical limiting activity has been observed in crystalline and amorphous materials ranging from bulk, thin films, monolayers, and quantum dots [1, 3, 5-9]. Typical examples are graphene oxide (GO), transition metal dichalcogenides, porphyrin, amorphous chalcogenide thin films, plasmonic materials, etc [1, 3, 5-11]. The primary nonlinear absorption processes that used for optical limiting are multi-photon and excited state absorptions (ESA) [1-4, 12-16]. In multi-photon absorption, the material simultaneously absorbs more than one photon since there is no real resonant single photon absorption state [15, 16]. This makes the process sensitive to high intensity. On the other hand, ESA is a two-step two photon absorption process in which the material first absorbs a photon and make the transition to the allowed resonant single photon state and in the second step, another photon is absorbed to make the transition to the next allowed single photon state [13, 16]. This step by step absorption of two photons through real states occurs in materials only when the excited state absorption cross-section is more than the ground state [16]. Typical examples of materials that show ESA are carbon-based materials such as fullerenes, phthalocyanine complexes, reduced graphene oxide (rGO), graphene oxide (GO) etc [1, 10, 13]. Among these, rGO(GO), a thin layer of graphite that contains non-uniform oxygen-containing functional groups are significant because it can surpass the zero-bandgap limitation of graphene [3, 12, 17, 18].

ESA of rGO(GO) is characterized by the presence of σ-states from $sp^3$ bonded carbons with oxygen [12, 18] that disrupts π-states from the $sp^2$ carbon sites [12, 18]. Enhancing the ESA of rGO(GO) nanosheets are of great importance to accomplish stronger and efficient light-matter interactions for applications like all-optical deterministic quantum logic and limiting at the nanoscale. Many efforts have been made to improve the ESA of rGO(GO) in the visible to IR region, for instance by modifying the electronic and chemical structure through functionalization and doping [1, 3, 5-8, 12-14]. Notwithstanding, the enhancements were minimal. Recent studies have shown that attaching metal (Ag/Au) and semiconducting (ZnO) nanoparticles onto GO can enhance the ESA [19-23]. For example, B. S. Kalanoor et al have demonstrated ESA in Ag nanoparticle functionalized graphene composite at high peak intensity of 13 GW/cm$^2$. Notably, below 13 GW/cm$^2$, these composites show only saturable absorption [23]. However, studies on Ag nanoparticle-GO nanocomposite revealed the enhancement by as much as 2.8 times at a peak intensity of 0.2 GW/cm$^2$ [21]. Despite observing small enhancements of ESA in Au/Ag-GO nanocomposites, however, the exact mechanism remains poorly defined. Further, the roles of ligand concentration in controlling the ESA of GO hybrids remain largely unexplored and needs to be resolved. Here we demonstrate experimentally and theoretically an extremely efficient scheme of ESA by charge transfer between AuNP and rGO(GO) as a new method to obtain unprecedented enhancement at low intensities that are not sufficient enough to significantly deplete the ground state of AuNP or rGO(GO). Strikingly, in AuNP-rGO hybrid structure, we find strong ESA coefficient ≈ 64±4 cm/GW at a moderate peak intensity of 0.16 GW/cm$^2$ that is in stark contrast to the saturable absorption of rGO and AuNP when excited at the plasmonic wavelength. However, the enhancements were much lower at other excitation wavelengths. This suggests that AuNP− rGO experiences high rates of charge transfer between them when excited at the plasmonic wavelength. Further, we could also observe the enhancement of ESA coefficient by one orders of magnitude at 1.6 GW/cm$^2$ in AuNP-rGO(GO) for plasmonic excitation. Remarkably, ESA coefficients of the hybrids increase with the increase in oxygen functional groups of GO. The unprecedented enhancements of the ESA are numerically simulated by using the five-levels charge transfer model. The charge transfer between the materials are experimentally verified by using the ultrafast pump-probe spectroscopy. Furthermore, the effects of donor-acceptor interactions have been analyzed at the AuNP-rGO(GO) interfaces by varying the functional ligand concentration with the help of first-principles calculations. Finally, we have fabricated liquid cell-based high-performance optical limiter with important device parameters better than that of the benchmark optical limiters.

## II. SAMPLE PREPARATION AND PHYSICAL CHARACTERIZATION

### A. Synthesis of gold nanoparticle (AuNP) on rGO(GO) nanosheets

We strategized the wet chemical synthesis based on the dispersible template method for synthesizing AuNP-rGO(GO) colloidal solutions and the detailed procedure is outlined in Fig. 1(a). In our method, the AuNP were synthesized in situ on the rGO(GO) surface by the standard low-cost citrate reduction technique (supplemental material (SM) [24]) [25, 26]. This procedure has advantages of high throughput and the flexibility of obtaining nanostructures with controllable size and morphology. The oxygen concentrations name the samples used in the present study as AuNP-GO, AuNP-rGO(400) and AuNP-rGO(1000), and the values inside the bracket show the temperature at which GO is heat treated. The oxygen percentage estimated from the CHN analysis are 61, 33 and 12 % for AuNP-GO, AuNP-rGO(400) and AuNP-rGO(1000) respectively (Table S1 in SM [24]).

### B. Morphological characterization and Raman analysis of AuNP-rGO(GO)

The powder X-ray diffraction (PXRD) pattern [27-29], scanning electron microscope (SEM) images and Raman spectra reveal that the samples show excellent crystallinity and the formation of spherical AuNP over rGO(GO) sheets. Figure 1(b) shows SEM images of GO which confirms nanosheet morphology. Figure 1(c) shows the SEM image of AuNP-GO, which reveal that the spherical AuNP were anchored over GO. Transmission electron microscope image of the AuNP shown in Fig. S1(a) of SM [24] indicate that the size of the nanoparticles was in the range of 25±3 nm with an overall spherical shape, together with a few larger particles of non-spherical shapes. SEM images of all other samples used in the present study are shown in Fig. S2 of SM [24]. "LabRam" high-resolution Raman spectrometer with the excitation source at 633 nm from a He-Ne laser in the confocal mode was used to record the Raman spectrum of the samples. Figure 2(a) shows the characteristic D mode at 1326 cm$^{-1}$ and G mode at 1580 cm$^{-1}$ of GO, rGO(400) and rGO(1000). The D mode is originating from the edges, defects, and disordered carbons, whereas the G mode is due to the phonons of sp$^2$ C atom of the graphite lattice [30, 31]. Apart from the two prominent D and G modes, we could also observe the weak 2D mode at 2645 cm$^{-1}$. The peak intensity ratio ($I_D/I_G$) of D and G modes depends on the degree of disorder and the graphitization of carbonaceous material [30, 31]. It can be seen from Fig. 2(a) that $I_D/I_G$ of GO, rGO(400) and rGO(1000) are 1.10, 0.86 and 0.83 respectively. Such an observation clearly indicates the removal of the oxygen-functional groups by our thermal exfoliation process which is also supported by CHN analysis (SM [24]). It is evident from the Fig. 2(b) that the hybrid samples also have the characteristic D, G and 2D modes at the same frequency of rGO(GO). However, the $I_D/I_G$ ratios show a slight increase as compared to rGO(GO) and the values are displayed in the Fig. 2(a & b). The increases in $I_D/I_G$ values of the hybrids indicate the nucleation of AuNP on the surfaces of rGO(GO) that introduce defects in the structure [30, 31].

**C. Linear optical response of AuNP-rGO(GO)**

To get a better understanding of the interaction between AuNP and rGO(GO) domains in the hybrid, we first recorded the optical absorption spectrum of AuNP, GO, rGO(400) and rGO(1000) and is shown in Fig. 2 (c). Optical absorption spectrum of AuNP is similar to the previous results that shows the plasmon absorption band at 530 nm which corresponds to the dipolar resonance between the localized plasmon oscillations and the excitation frequency [32]. In contrast, GO exhibit a broad absorption, with two absorption peaks, one at 231 nm due to $\pi$ to $\pi^*$ transitions in aromatic C=C bonds and the other at 303 nm from $n$ to $\pi^*$ transition of carbonyl (C=O) group [8, 30]. It can be seen in the Fig. 2(c) that rGO(400) and rGO(1000) have broad absorption from 200 to 1100 nm with the optical bandgap in the infrared region since oxygen-containing groups forming the $sp^3$ matrix are converted into $sp^2$ clusters, which are in agreement with the published reports [8, 30, 31]. Optical absorption spectrum of AuNP-rGO(GO) shown in Fig. 2(d) is exactly similar to the rGO(GO) presented in the Fig. 2(c). However, the band corresponding to the plasmonic effect of AuNP is absent, which we assume that the strong absorption background of rGO(GO) hide the plasmon band since the AuNP loading on rGO(GO) is less.

### III. RESULTS AND DISCUSSION

#### A. Nonlinear optical response of AuNP-rGO(GO)

A standard open aperture Z-scan technique was used to measure the nonlinear optical response of AuNP-rGO(GO) hybrid structures, which measures the total transmittance as a function of incident laser intensity [16, 33, 34]. The excitation wavelengths in our experiments are at 1064 and 532 nm, 5 ns pulses from the fundamental and second harmonics of the Nd: YAG laser. The repetition rate of the laser was set at 10 Hz to provide enough time for the system to relax after each pulse. The beam is focused along the Z-axis of a computer controlled translation stage, using a 200 mm focal length lens. The Rayleigh length ($Z_0$) and the beam waist of our experiments are 3.2 mm (1064 nm), 1.6 mm (532 nm), and ~ 32 (1064 nm), 17μm (532 nm) respectively. For Z-scan measurements, we have prepared the dispersions in the distilled water, with a linear transmittance (T) of ≈ 70%.

Figure 3(a) shows the Z-scan peak shape of AuNP-rGO(1000), AuNP, and rGO(1000) at an on-axis peak intensity (measured at the focal point) of 0.16 GW/cm$^2$ for 532 nm (plasmon wavelength) excitation. The normalized transmittance of AuNP-rGO(1000) shows unprecedented nonlinear absorption that is in stark contrast to the saturable absorption of AuNP and rGO(1000). At this moderate intensity, the pristine rGO(1000) and AuNP show saturable absorption since the materials have resonant single photon absorption states and the peak intensity is not sufficient to make the excited state absorption cross-section more than the ground state. Therefore, we assume that the unprecedented nonlinear absorption of the hybrid is due to the ESA invoked by the charge transfer between the individual components which was verified later. The Z-scan theory discussed in the SM [24] was used to fit the data to derive the ESA coefficient (β) and saturation intensity (Is). From the best fit to the normalized transmittance, we found the $β_{532\ nm}$ of AuNP-rGO(1000) ≈ 64 ±4 cm/GW. Next, we have selected the hybrid samples of AuNP-rGO(400) and AuNP-GO, where rGO(400) and GO show ESA. As can be seen in the Fig. 3(b) and Table 1 that the $β_{532\ nm}$ of AuNP-rGO(400) ≈ 112±3 cm/GW, which is found to be ≈ 6 times of the rGO(400). Likewise, $β_{532\ nm}$ of AuNP-GO (130±6 cm/GW) is ≈ 6 times of GO (Fig. 3(c)). Figure 3(d) shows the normalized transmittance as the function of input intensity at a peak intensity of 1.6 GW/cm$^2$ for 532 nm pulse excitation. The nonlinear optical properties of the samples are summarized in Table. 1. To get more insight on ESA of the AuNP-rGO(GO) hybrid samples, we have performed intensity dependent Z-scan experiments. As expected from the theory of ESA, $β_{532\ nm}$ of the hybrids show strong intensity dependence (see Table. 2). Strikingly, we could observe enhancements of the ESA by as much as an order of magnitude. For instance, enhancement in AuNP-rGO(GO) is 6 times at 0.16 GW/cm$^2$ and 10 times at 1.6 GW/cm$^2$. At this stage, we assume that the strong ESA of AuNP-rGO(GO) is due to the charge transfer between AuNP and rGO(GO) that can be used as a new method to enhance the ESA even in materials that show only saturable absorption at moderate intensities.

Before explaining the ESA of the hybrid, we have carefully analyzed the results of AuNP and rGO(GO). The normalized transmittance of AuNP shows saturable absorption which is arising from the depletion of the ground-state surface plasmons since the rate of excitation is much faster than the rate of relaxation to the ground state [35]. Saturable absorption of rGO(1000) is ascribed to the Pauli blocking since the excited state becomes almost occupied, and the Pauli exclusion principle prevents further absorption [8]. Next, we have analyzed the normalized Z-scan peak shape of rGO(400) and GO, which shows a decrease in transmittance as a function of input intensity. This explains an ESA channel is introduced at high intensities that are connected to the sp$^3$ hybrid states containing oxygen functional groups [8].

### B. Charge Transfer between donor-acceptor materials

At this point, it is essential to confirm that the observed enhancement in ESA is due to the synergistic effect of charge transfer between AuNP and rGO (400)/GO. As the first approximation, if rGO(400)/GO and AuNP are independently responding to the incoming light, then we should have expected a reduction in the ESA. For the range of intensities when $I < I_s$ Eq. S2 in SM [24] can be approximated by Taylor expansion to

$$\alpha(I) \approx \alpha_0 + \left(\beta - \frac{\alpha_0}{I_s}\right) I \tag{1}$$

where $\alpha(I)$ and $\alpha_0$ are intensity dependent absorption coefficient and linear absorption coefficient respectively. From Eq. 1 it can be seen that $\alpha(I)$ decreases with increase in intensity.

However, in our experiment (Fig. 3(a-c)), we have observed an increase in α(I) of the hybrid. Therefore, we come to a logical conclusion that the carriers excited in AuNP through linear absorption, which led to saturable absorption, should have contributed to the ESA in the hybrid.

To confirm the enhancement of ESA is due to charge transfer, we have performed the Z-scan experiment at 1064 nm, which is far away from the resonant absorption of AuNP. Excitation of the sample at this wavelength cannot promote the electrons from the Fermi level of AuNP to rGO(GO). In this scenario, it is expected that the ESA of the hybrid is same as that of rGO(GO) due to the absence of charge transfer. Fig. 3(e & f) presents the Z-scan traces of the hybrid samples. As expected, we have observed that the ESA of AuNP-rGO(GO) and rGO(GO) are nearly the same within the error bars. These results certainly ascertain our charge transfer model and demonstrate that strong ESA can be achieved when the hybrid architecture can potentially facilitate light-induced charge transfer. Our results are important in nonlinear optics, where we can customize ESA and design hybrid materials for optical limiting applications.

### C. Ultrafast time resolved pump-probe spectroscopy of AuNP-GO hybrid

Most remarkably, our experimental findings suggest that the enhancement of the ESA of the hybrid is primarily due to charge transfer and therefore to shed more light on this mechanism, we have performed broadband transient absorption (TA). In this experiment, the samples were excited with 400 nm, 120 fs pulses of fluence 300 μJ/cm$^2$, and the change in absorption of the sample ($\Delta A=-(\log[I_{ex}/I_0])$) is probed by ultrashort white light pulses in the wavelength range 450–650 nm. Here $I_{ex}$ and $I_0$ were the transmitted intensities of the sequential probe pulses after a delay time τ following excitation by the pump beam, and in the ground state, respectively. Kinetic traces were obtained by varying the path lengths of the probe beam using a micrometer resolution translation stage and chirp corrected using the procedure reported elsewhere [36]. To avoid the heating effects, the sample was rotated continuously.

TA spectrum of AuNP-rGO shows a spectrally flat response which is not helpful in analyzing the charge transfer. Figure 4(a) presents the TA spectrum of AuNP-GO, AuNP, and GO at a pump-probe delay of 1 ps. TA of AuNP and GO are dominated by bleach at the plasmon resonance (530 nm) [37-39] and a broad ESA [8, 40] respectively. Detailed analysis of the spectra can be found in SM [24]. Strikingly, TA of the hybrid sample has different spectral shape and distinct time-dependent dynamics (Fig. 4(b)). First, spectrum of the hybrid is not a simple addition of AuNP and GO at any time within the timescale of our experimental window. Second, the new bleach signal below 500 nm shows a redshift with the increase in pump-probe delays. Both features encode information of the net charge transfer between AuNP and GO. In the next step to extract the charge transfer, we have plotted the integrated area of 525-540 nm of the plasmonic region (to remove the effect of redshift)) as a function of time and the resulting kinetic traces are shown in Fig. 4(c-d). It can be seen from the figure that the TA of AuNP-GO is faster than AuNP and GO. To calculate the charge transfer, we assumed that the dynamics of TA in the plasmonic region show a difference between the hybrid and AuNP, then the rate of charge transfer (k) can be estimated as $1/\tau_{(AuNP-GO)} - 1/\tau_{(AuNP)}$ [41, 42]. Here $\tau$ is the average lifetimes of the decay (see in the SM [24]). By this method, we have found the value of k = (4.3 ±0.5)× $10^{11}$ $s^{-1}$.

**D. Numerical simulations of five level rate equations through charge transfer**

After demonstrating the hallmark observation of the unprecedented enhancement of the ESA due to charge transfer, we have developed five-level rate equation model shown in Fig. 4(e) to simulate the results numerically. Here $D_1$ and $D_2$ stand for the ground and excited levels respectively of the donor. Similarly, $A_1$, $A_2$, and $A_3$ represent the ground, first excited and second excited levels respectively of the acceptor. In our model, upon laser illumination, donor AuNP undergoes one-photon absorption and transfer the excited electron to the first excited state $A_2$ of the acceptor rGO(GO). From the first excited state of the acceptor, it absorbs another photon to go to the second excited state. Such a charge transfer results in the enhancement of ESA and can be represented by the following coupled rate equations

$$\frac{dN_{D1}}{dt} = -\frac{\sigma_0 I\, N_{D1}}{\hbar\omega_0} + \frac{N_{D2}}{\tau_S} \tag{2}$$

$$\frac{dN_{D2}}{dt} = \frac{\sigma_0 I\, N_{D1}}{\hbar\omega_0} - \gamma_c N_{D2} - \frac{N_{D2}}{\tau_S} \tag{3}$$

$$\frac{dN_{A1}}{dt} = -\frac{\sigma_1 I\, N_{A1}}{\hbar\omega_0} + \frac{N_{A2}}{\tau_2} \tag{4}$$

$$\frac{DN_{A2}}{dt} = \frac{\sigma_1 I\, N_{A1}}{\hbar\omega_0} + \gamma_c N_{D2} + \frac{N_{A3}}{\tau_1} - \frac{N_{A2}}{\tau_2} - \frac{\sigma_{ES} I\, N_{A2}}{\hbar\omega_0} \tag{5}$$

$$\frac{dN_{A3}}{dt} = \frac{\sigma_{ES} I\, N_{A2}}{\hbar\omega_0} - \frac{N_{A3}}{\tau_1} \tag{6}$$

Here I is incident laser intensity, and subscripts in N ($N_{D1}$, $N_{D2}$, $N_{A1}$, $N_{A2}$, $N_{A3}$) represent the respective donor (D) and acceptor (A) levels. The lifetimes $\tau_1$ = 0.963 ps, $\tau_2$ = 13.08 ps and $\tau_s$ = 5.17 ps were obtained from the TA. Absorption cross-sections calculated from the Z-scan data [43, 44] are listed in Table. 3. Charge transfer rate ($\gamma_c$) was estimated as (4.3 ±0.5) × $10^{11}$ s$^{-1}$ from the TA. The solutions of Eqs. (1)–(6) were obtained numerically by using a Gaussian pulse both in the temporal and spatial domains. The transmitted intensity through the sample is given by

$$\frac{dI(Z)}{dZ} = [-\sigma_0 N_{D1} - \sigma_1 N_{A1} - \sigma_{ES} N_{A2}]\, I(Z) \tag{7}$$

Figure 4(f) shows the numerically simulated normalized transmitted intensity through the sample as a function of the position. It can be seen from the figure that our numerical simulations accurately, reproduce the experimental results. Therefore, we conclude that the enhancement of the ESA of AuNP-rGO(GO) is due to the synergistic charge transfer between AuNP and rGO(GO) domains, which can be consistently described by a five-level charge transfer model. Further, we have endeavored to unravel the charge-transfer mechanism at AuNP-rGO(GO) interface by scrutinizing their electronic structure.

**E. Charge transfer between AuNP-rGO(GO) interface from first-principles**

Although simulating the actual experimental structures of rGO(GO) is extremely complicated, we have adopted a simpler scheme to extrapolate the experimental results. To perceive the significance of functional groups in mediating the interlayer coupling of AuNP-GO, we have constructed five control systems, *viz*., (1) Graphene nanosheet (G), (2) G with a single hydroxyl (OH) group (G-OH), (3) G with a single carboxylic (COOH) group (G-COOH), (4) G with both OH and COOH groups in lower concentration (GO(LOW)) and (5) G with both of these groups in higher concentration (GO(HIGH)). Among these systems, GO(LOW) and GO(HIGH) are approximate representative of rGO and GO respectively. All of these systems were investigated with the help of ab-initio Density-functional theory (DFT) based formalism using plane-wave pseudopotential approach with projector augmented wave (PAW) potentials as implemented in Vienna Ab-initio Simulation Package (VASP) [45, 46]. Ionic and lattice parameter optimization of the constructed surface and interfaces are obtained by conjugate gradient algorithm until the Hellmann-Feynmann forces on each ion is less than 0.01eV/Å. For self-consistent calculations and structure optimization, an energy cutoff of 500 eV is used with a *k*-point mesh size 5×5×3 with an energy accuracy of 0.0001 eV. Calculation of electronic charge densities for optimized structures, obtained from VASP calculations, is carried out using double-zeta plus polarization basis set, as implemented in ATK 15.1 package [47, 48]. Electron correlation within the system is treated using Perdew-Burke-Ernzerhoff exchange-correlation functional under spin-polarized generalized-gradient approximation (GGA) for all calculations. To account for the interface-induced dipolar interaction, we have incorporated the van der Waal corrections by using Grimme DFT-D3 method [49]. The details of the theoretical calculation and the control systems (1-3) can be found in SM [24].

Fig. 5 (a-c) shows the atom projected density of states (APDOS) and orbital projected density of states (OPDOS) of Au-Graphene(G), Au-GO(LOW) and Au-GO(HIGH) interfaces. The observed general trend after comparing the DOS of the control systems (1) to (5), as presented in SM [24], and the corresponding interface with Au, indicates that the originally semiconducting systems with varying bandgap turn into metallic ones after combining with Au layer. The only exception was the system (5), where in addition to strong spin-polarization, there is a very small bandgap. From the Fig. 5(a), we can see that there is charge transfer from the C-$2p$ levels of graphene sheet to Au $6s$ levels, rendering a shift of Fermi-level of the combined system ($E_F$) towards valence band by ~ 0.2 eV, which implies a $p$-type doping of the G sheet. Similar is the case with single ligands (G-OH and G-COOH), where the charge transfer occurs from G-OH and G-COOH to Au, with the shift of $E_F$ towards valence band by ~ 0.2 and 0.1 eV respectively (SM [24]). However, with increasing ligand concentration, the $E_F$ shifts towards conduction band by ~ 0.04 and 0.6 eV respectively in Au-GO(LOW) and Au-GO(HIGH), as in Fig. 5 (b) and (c). At higher ligand concentration, the structural distortion due to the $sp^2$-$sp^3$ conversion brings the Au-layer closer leading to a charge transfer from Au-$6s$ (donor) to C-$2p$ (acceptor) via the functional ligands, amounting to $n$-type doping of the corresponding GO sheet. Respective charge-density profile and the Z direction projection of the electronic charge density are plotted in Fig. 5 (d-e). First and the second peaks of the Z direction projection of the electronic charge densities correspond to the GO(G) and the Au layer respectively. The increase of charge density at the intermediate region and the shift of the peaks indicate the increasing charge transfer with increasing ligand concentration. The quantified shift of $E_F$ all control systems are listed in Table 4. The presence of higher concentration of ligands for Au-GO(LOW) and Au-GO(HIGH) introduces asymmetry in the charge-transfer and thus are more spin-polarized. Thus, first-principles studies demonstrate that the donor-acceptor configurations and indicate the direction of charge-transfer in the AuNP-rGO(GO) systems pointing out its dependence on the concentration and type of functional ligands attached to the GO sheet. For AuNP-GO interface with different ligand concentrations, it is indeed possible to modulate the interlayer coupling between the two layers and thereby to control the structural distortion and charge transfer.

### F. Liquid cell based optical limiter

Clearly, the strong ESA presented here are very large; however, their usefulness would be rather limited without showing a device fabricated based on this idea. To do so, we have fabricated a liquid based optical limiter and the schematic of our device is shown in Fig. 6(a). It essentially consists of a 1 mm path length liquid cell, that contains the aqueous solution of AuNP-rGO(GO) as optical limiting media. We have prepared the limiting medium in distilled water, with a linear transmittance (T) of $\approx$ 70%. The device was tested using the Z-scan setup since the optical limiting relies on the fact that the transmission decreases when the input laser intensity is above a threshold. That means, the optical limiting device is designed to keep the linear transmittance below some specified maximum value of input intensity and limit above a certain threshold intensity. To evaluate the device performance, it is essential to characterize the strength of optical limiting using onset threshold $F_{ON}$ (the input intensity at which normalized transmittance starts to deviate from linearity), optical limiting threshold $F_{OL}$ (the input intensity at which normalized transmittance drops below 50%) and limiting differential transmittance $T'_c$ ($dI_{out}/dI_{in}$) at high intensity.

To demonstrate the optical limiting of our device, we have measured the transmitted intensity ($I_{out}$) as a function of the input intensity ($I_{in}$) of the laser beam. Fig. 6(b) presents a plot of output intensity versus input intensity of our device measured for the 5 ns pulses of wavelength 532 nm upto a peak intensity of 0.78 GW/cm$^2$. Our measurements clearly indicate that $I_{out}$ through the device is linearly proportional to $I_{in}$ at lower intensity. However, with increase in $I_{in}$, a stage reaches where the $I_{out}$ is no longer linearly proportional to $I_{in}$, i.e. deviation of the curve from the linear transmittance. To calculate $F_{ON}$, $F_{OL}$ and $T'_C$, we have shown in Fig. 6(c) the normalized transmittance as a function of the input intensity and the estimated values are summarized in Table 5. Notably, an ideal optical limiter should have very low $F_{ON}$, $F_{OL}$, and $T'_C$. From the table, it can be seen that these important device parameters of are better than the benchmark optical limiters such as graphene oxide nanosheets, MoS$_2$, graphene nanoribbon, CdS nanoparticle and Pd nanowire [50-54] (Table 5), which makes our device as an exceptionally good optical limiter. Further, the advantage of our liquid based optical limiter is that they self-heal, permitting high dynamic ranges limited only by damage to cell windows. In our device, dynamic range is estimated to be 120 GW/cm$^2$.

## IV. CONCLUSIONS

Experimental and theoretical results have shown that our new scheme of ESA by efficient charge transfer of a donor-acceptor material as a new method to enhance the nonlinear absorption in AuNP-rGO(GO) hybrid materials. At this point, we envision that our current work represents a significant advance in nonlinear optics and opens up new avenues, ranging from fundamental investigations to technology applications. For example, fundamental perspective of our study is that the charge transfer occurs from AuNP to GO at higher ligand concentration, and the situation gets reversed at low ligand concentration, which is not known before. The excellent agreement between experimental and theoretical results in our work, provide test beds for design guidelines to make hybrid structures for enhancing, tuning and modulating the nonlinear optical response. Additionally, we have fabricated an optical limiting device based on liquid cell with important device parameters better than that of the benchmark optical limiters.

ACKNOWLEDGMENTS

The authors gratefully acknowledge the Department of Science and Technology (Project no: EMR/2016/002520), DAE BRNS (Sanction no: 37(3)/14/26/2016-BRNS/37245) and. A.K.M. thanks INSPIRE faculty grant (Grant No. IFA14-MS-25). T.M. and D.K. acknowledge the BARC ANUPAM supercomputing facility.

**Table 1.** Excited state absorption coefficient (β) and saturation intensity ($I_s$) calculated by fitting the experimental data with equations shown in SM [24].

| Sample | $\beta_{532\ nm}$ (cm/GW) | $\beta_{1064\ nm}$ (cm/GW) | $I_{sat\ 532\ nm}$ (GW/cm$^2$) |
|---|---|---|---|
| AuNP | --------- | --------- | 0.05 ± 0.01 |
| rGO(1000) | --------- | 39±5 | 0.06 ± 0.01 |
| rGO(400) | 13±2 | 46±4 | --------- |
| GO | 18 ± 2 | 62±6 | --------- |
| AuNP-rGO(1000) | 64 ±4 | 34±4 | --------- |
| AuNP-rGO(400) | 112 ± 3 | 40±5 | --------- |
| AuNP-GO | 130 ± 6 | 59±6 | --------- |

**Table 2.** Comparison of excited state absorption coefficient of AuNP-rGO(GO) hybrid at different peak intensities for 532 nm excitation.

| Sample | $\beta_{532\ nm}$ (cm/GW) | | |
|---|---|---|---|
| | 0.16 GW/cm$^2$ | 0.78 GW/cm$^2$ | 1.60 GW/cm$^2$ |
| AuNP-rGO(1000) | 64±4 | 115±4 | 159±6 |
| AuNP-rGO(400) | 112±3 | 155±5 | 184±5 |
| AuNP-GO | 130±6 | 162±5 | 187±6 |
| rGO(1000) | ------- | 5±2 | 8±1 |
| rGO(400) | 13±2 | 14±2 | 16±2 |
| GO | 18±2 | 19±2 | 20±2 |

**Table 3.** Ground state ($\sigma_0$, $\sigma_1$) and excited state ($\sigma_{ES}$) absorption cross sections of AuNP-rGO(GO) for λ= 532 nm.

| Sample | ($*10^{-18}$ cm$^2$) | | $\sigma_{ES}$ ($*10^{-18}$ cm$^2$) |
|---|---|---|---|
| | $\sigma_0$ | $\sigma_1$ | |
| AuNP | 8.0 | ----- | ----- |
| rGO(1000) | ----- | 0.13 | 0.08 |
| rGO(400) | ----- | 0.14 | 0.16 |
| GO | ----- | 0.15 | 0.19 |
| AuNP-rGO(1000) | ----- | 0.59 | 0.81 |
| AuNP-rGO(400) | ----- | 0.60 | 0.96 |
| AuNP-GO | ----- | 0.61 | 1.16 |

**Table. 4:** Shift in Fermi-levels ($E_F$) in comparison to G-sheet ($\Delta E_{F-G}$) and GO ($\Delta E_{F-GO}$).

| System | $\Delta E_{F-G}$ | $\Delta E_{F-GO}$ |
|---|---|---|
| Graphene(G) | ----- | ----- |
| GO(LOW) | -0.33 | ----- |
| GO(HIGH) | -0.82 | ----- |
| Au-G | -0.19 | -0.19 |
| Au-GO(LOW) | -0.29 | +0.04 |
| Au-GO(HIGH) | -0.15 | +0.67 |

**Table. 5:** Comparison of nonlinear optical parameters such as $F_{ON}$, $F_{OL}$ and $T'_C$ of different Optical limiters at 532 nm.

| System | Lin T | $F_{ON}$ (J/cm$^2$) | $F_{OL}$ (J/cm$^2$) | $T'_C$ | References |
|---|---|---|---|---|---|
| AuNP-GO | 70 | 0.05 | 0.49 | 0.18 | *Present work |
| AuNP-rGO(400) | 70 | 0.05 | 0.70 | 0.24 | *Present work |
| AuNP-rGO(1000) | 70 | 0.06 | 1.09 | 0.28 | *Present work |
| GO | 70 | 0.19 | 1.19 | ---- | [39] |
| MoS$_2$ nanosheets | 63.8 | 1.52 | 11.16 | ---- | [40] |
| Graphene nanoribbon | 70 | 0.10 | 0.70 | ---- | [41] |
| CdS nanoparticles | 70 | 0.30 | 2.55 | 0.56 | [42] |
| Pd nanowire | 80 | 0.09 | 0.90 | ---- | [43] |
| Graphene thin film | 73 | 0.01 | 0.08 | 0.05 | [1] |

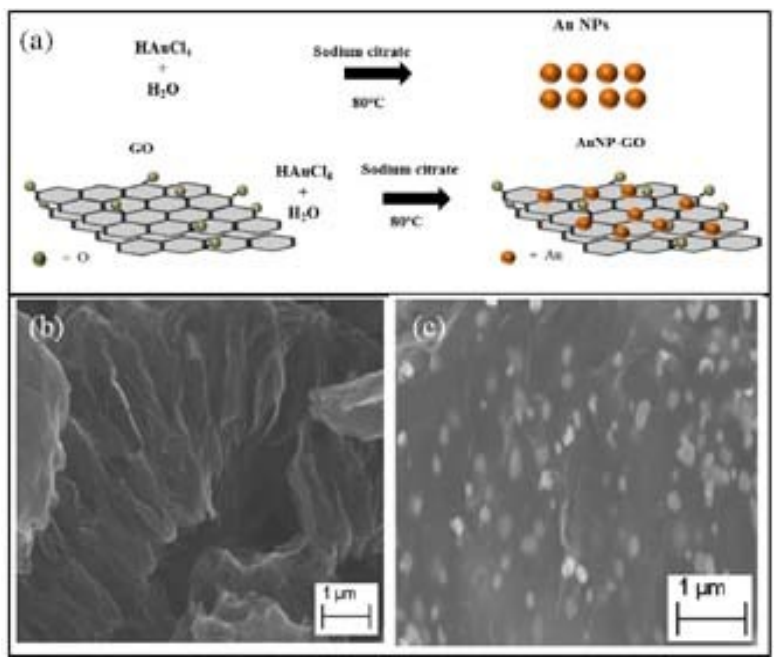

**Fig. 1.** (a) Schematic representing synthesis of AuNP and AuNP-GO hybrid, (b) and (c) SEM images of GO and AuNP-GO hybrid respectively. The images reveal nanosheet like morphology of few layer GO and formation of spherical AuNP over GO nanosheet.

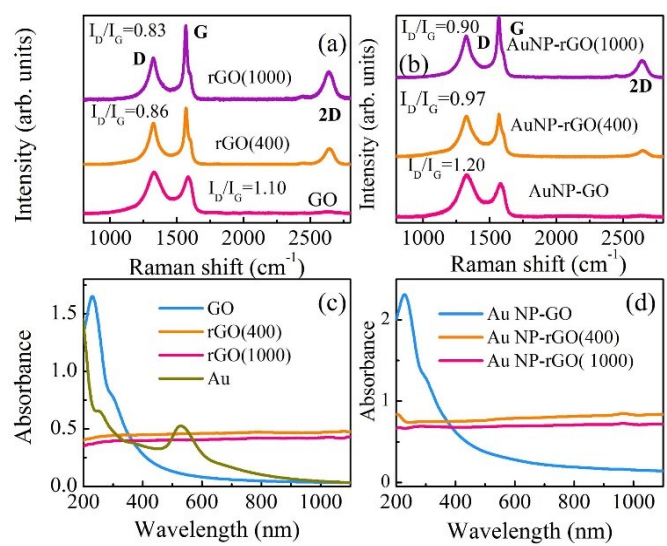

**Fig. 2.** Raman spectrum of (a) rGO(1000), rGO(400) and GO and (b) AuNP-rGO(1000), AuNP-rGO(400) and AuNP-GO. Optical absorption spectrum of (c) rGO(1000), rGO(400), GO and AuNP and (d) AuNP-rGO(1000), AuNP-rGO(400) and AuNP-GO. We assume that the strong absorption background of GO and rGO hide the plasmon absorption of AuNP, where the AuNP loading on rGO(GO) is very less.

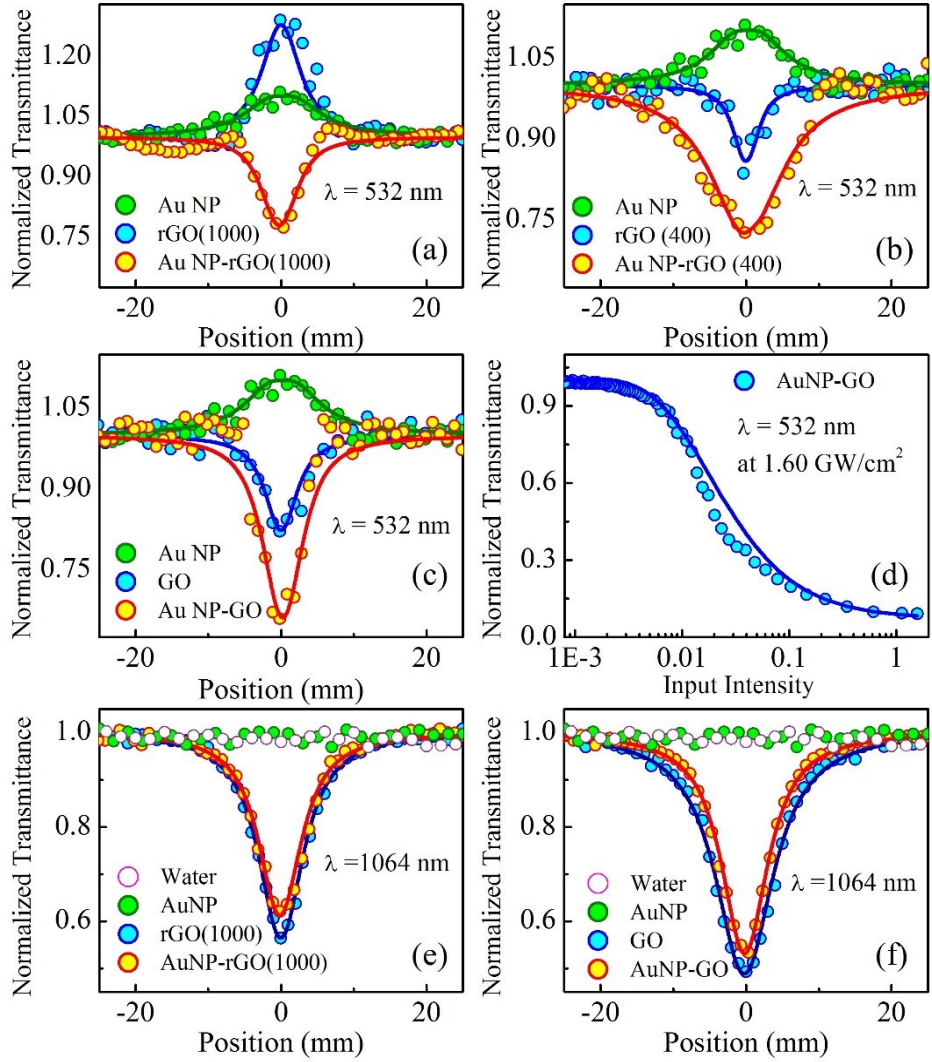

**Fig. 3**. (a-c) Normalized transmittance as a function of position in open-aperture Z scan at 5ns, 532 nm pulse excitation at a peak intensity of 0.16 GW/cm$^2$, for the samples indicated in the panel. (d) Normalized Transmittance as a function of input intensity for AuNP-GO at input intensity 1.6 GW/cm$^2$. (e & f) Normalized transmittance as a function of position in open-aperture Z scan at 5ns, 1064 nm pulse excitation at a peak intensity of 0.35 GW/cm$^2$. It is evident from Fig. 3(e & f) that distilled water has no nonlinear optical response. The solid lines in the figure show the theoretical fitting.

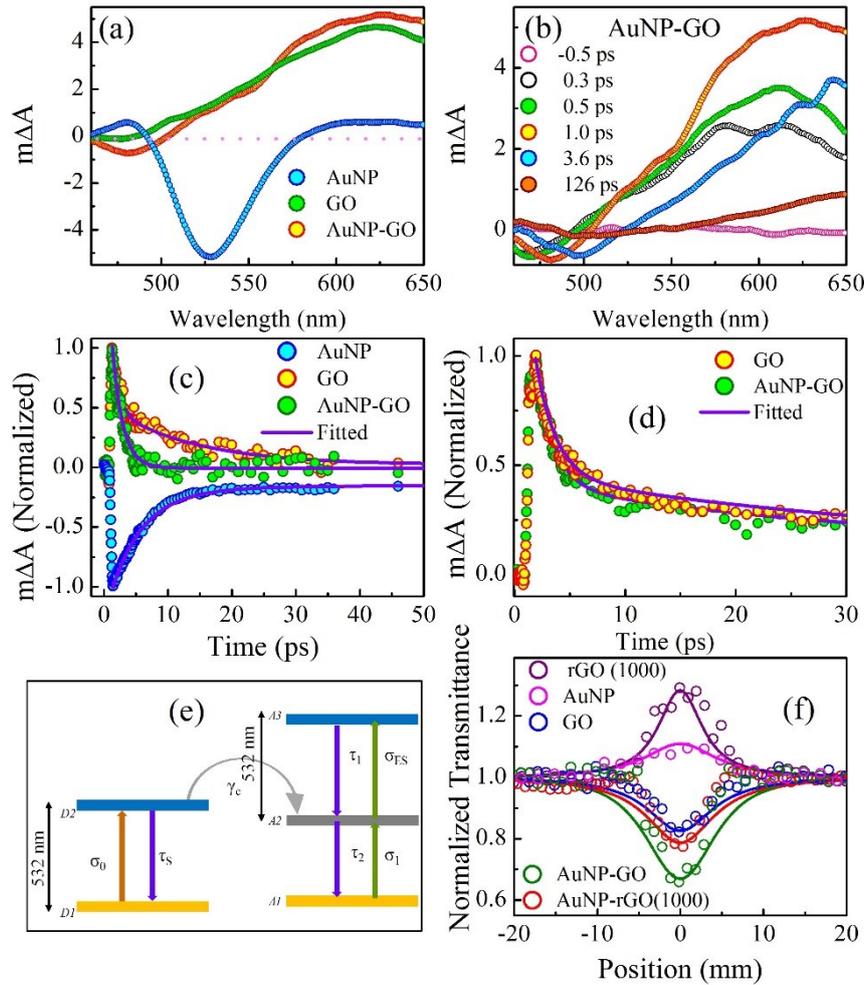

**Fig. 4.** (a) Spectral evolution of TA spectra of GO, AuNP and AuNP-GO hybrid at 1 ps, (b) TA spectra at selected pump-probe delays of AuNP-GO hybrid, as a function of wavelength, (c) Temporal evolution of TA spectra of GO, AuNP and AuNP-GO hybrid integrated over the region 525-540 nm, TA spectra are faster in the case of hybrid compared to GO and AuNP, indicates charge transfer through donor- acceptor intraction. (d) Temporal evolution of TA spectra of GO, AuNP and AuNP-GO hybrid at region 600-650 nm, (e) Schematic diagram of charge transfer through donor-acceptor interaction in a 5-level model. Here $D_1$ and $D_2$ stand for the ground and excited levels respectively of the donor. Similarly, $A_1$, $A_2$, and $A_3$ represent the ground, first excited and second excited levels respectively of the acceptor. (f) Shows numerically simulated data which matches with the experimental result.

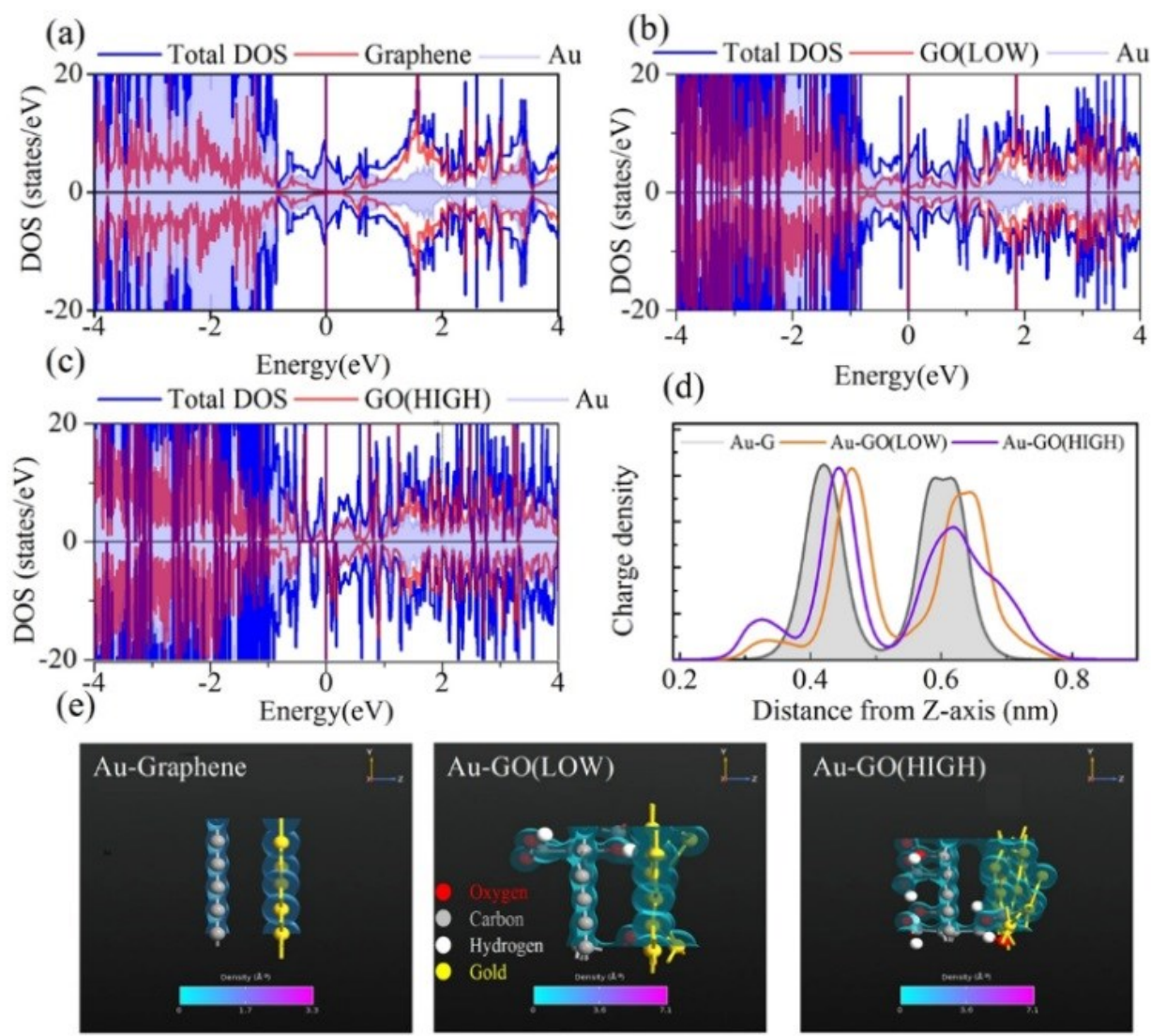

**Fig. 5.** The atom projected density of states (APDOS) of (a) Au-G (b) Au-GO(LOW) (c) Au-GO(HIGH). All the DOS plots are with respect to ($E - E_F$). The zero of energy axis is the Fermi-level. (d) One-dimensional and (e) three-dimensional charge density projection of Au-G, Au-GO(LOW), Au-GO(HIGH).

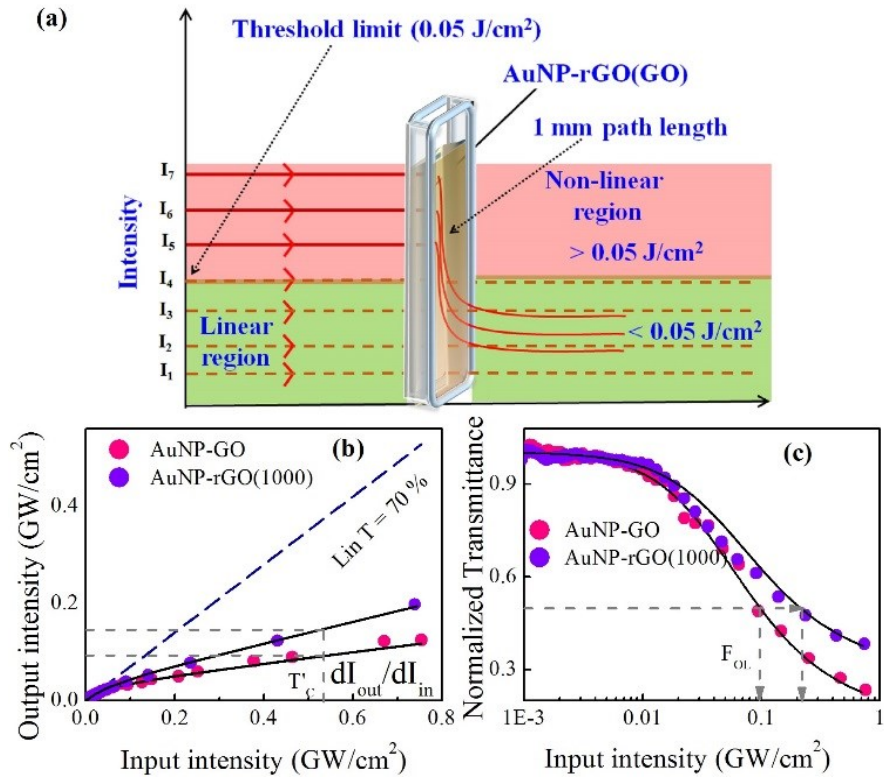

**Fig. 6.** (a) Schematic of our liquid based optical limiting device. It essentially consists of a 1 mm path length liquid cell, that contains aqueous solution of AuNP-rGO(GO) acting as optical limiting media. The concentration of the limiting medium in the distilled water is 0.5 mg/ml, with linear transmittance (T) ≈ 70%. When the lower intensity beams falls in the linear and safety region (green region), our device will allow the light to pass through. However, above the threshold limit (red region), the system will act as an optical limiter and attenuate the intense beam. (b) Output intensity as a function of input intensity for 532 nm pulse excitation. The dashed and solid lines in the figure show the linear transmittance (T) and the theoretical fitting, respectively. (c) Normalized Transmittance as a function of input intensity.